\documentstyle[aps,amsfonts]{revtex}
\newcommand{\g}{\hbox{\sl g}}
\newcommand{\D}{\hbox{$\delta$}}
\newcommand{\DD}[2]{\frac{\partial #1}{\partial #2}}
\newcommand{\dd}{\text{d}}
\newcommand{\B}[1]{\hbox{$\protect\bbox{{\Bbb #1}}$}}
\newcommand{\BT}[1]{\hbox{$\widetilde{\protect\bbox{{\Bbb #1}}}$}}
\newlength{\LENs}
\newlength{\LENn}
\newlength{\LENl}
\newlength{\LENL}
\def\LENset{\ifpreprintsty \LENs=2.8mm \LENn=3.0mm \LENl=4.5mm \LENL=5mm%
\else \LENs=1.16mm \LENn=1.9mm \LENl=3.4mm \LENL=3.8mm \fi}
\LENset
\newcommand{\m}[1]{\makebox[\LENs][c]{\raisebox{0.3mm}{-}$#1$}}
\newcommand{\p}[1]{\makebox[\LENs][c]{$#1$}}
\newcommand{\mL}[1]{\makebox[\LENl][c]{-$#1$}}
\newcommand{\pL}[1]{\makebox[\LENl][c]{$#1$}}
\newcommand{\tld}[1]{\hbox{$\vphantom{#1}\smash{\tilde{#1}}$}}
\newcommand{\eref}[1]{\hbox{(\ref{#1})}}
\newcommand{\mass}{\text{\sl m}}
\newcommand{\mc}{\mass}

\newcommand{\Cdot}{\,}
\newcommand{\SC}{{\frak s}}
\newcommand{\II}{{\frak I}}
\newcommand{\AAA}{{\cal A}}
\newcommand{\KK}{{\cal K}}
\newcommand{\CC}{C}
\newcommand{\W}{{\rm\scriptscriptstyle W}}
\newcommand{\RR}{{}_{\rm\scriptscriptstyle R}}
\newcommand{\LL}{{}_{\rm\scriptscriptstyle L}}
\begin{document}
\draft
\widetext

\title{Generalized equation of relativistic quantum mechanics in a gauge
field}

\author {A.~A.~Ketsaris\thanks {E-Mail: {\tt ketsaris@sai.msu.su}}}
\address{19-1-83, ul. Krasniy Kazanetz, Moscow 111395, Russian Federation}
\date {September, 1999}
\maketitle

\begin{abstract}
We develop an unified algebraic approach to the description of gauge
interactions within the framework of a new concept of quantum mechanics.
The next step in generalizing the space-time and the action vector space is
made. The gauge field is defined through linear mappings in the generalized
space-time and the action space.  Relativistic quantum mechanics equations
for particles in a gauge field {\it are derived\/} from the structure equations
for the action space expanded in the linear mappings of action vectors.  In
a special case, these equations are reduced to the relativistic equations for
the leptons in the electroweak field. As against the standard
Glashow--Weinberg--Salam model,
the set of equations includes the equation for the right neutrino interacting
only with the weak $Z$-field.
\end{abstract}

\section{Introduction}

The new concept of quantum mechanics has been put forward in
our previous works \cite{book,art1}. The main features of this concept are
the following. We introduced the space of all contravariant tensors over the
usual space-time as {\it a generalized space-time\/} $\B{X}$.  The space of the
Clifford algebra $\B{C}_4$, selected from the generalized space-time $\B{X}$,
was used for the description of leptons and hence was called {\it a space of
leptons}.  The action was considered as a vector quantity.  The action
vectors formed an algebra $\B{SX}$ similar to the algebra $\B{X}$.  The wave
function was identified with a differential of action vector. Relativistic
quantum mechanics equations for free particles were derived from the structure
equations for the algebra $\B{SX}$.

In the present work, we develop the specified concept of quantum mechanics
for the purpose of describing the interaction of particles with a gauge field.
The following propositions are used as the basis for our study:
\begin{enumerate}
\item
Linear mappings of the generalized space-time $\B{X}$ onto themselves are
introduced.  The linear mappings form algebra $\B{U}$. A kinematic
space $\B{T}=\B{X}+\B{U}$, endowed with algebraic properties, is defined.

\item
The action is considered as vector quantity. The action vectors
form an algebra $\B{S}$ similar to the algebra $\B{T}$.

\item
A gauge potential is defined as a derivative of linear mapping coordinates
in the generalized space-time and the space of generalized action.

\item
The partial derivation of the multiplication rule for the algebras $\B{T}$ and
$\B{S}$ result in the specific differential relations called the structure
equations. The structure equations for algebra $\B{S}$ are reduced to the
generalized equations of quantum mechanics for particles in a gauge field.

\end{enumerate}

\section{Linear mappings of the generalized space-time}

\subsection{Vector space of linear mappings}

We consider {\it linear mappings\/} of the generalized space-time $\B{X}$. Let
us introduce a linear operator $h(\ )$ mapping $\B{X}$ to itself:
\[
     x' = h(x) \,, \qquad\qquad x,x'\in \B{X}\,.
\]
The vector of the generalized space-time can be decomposed along
basis vectors $e_I$:
\[
       x = e_0\Cdot x^0 + e_{i_1}\Cdot x^{i_1} +
           e_{i_2 i_1}\Cdot x^{i_1 i_2} + \ldots +
           e_{i_n \ldots i_2i_1}\Cdot x^{i_1 i_2\ldots i_n} + \ldots
         = e_I\cdot x^I  \,,
\]
where $e_0$ is the unit of reals, $\{e_{i_1}\}$ is the basis in the usual
space-time, the lower collective index
\[
       I= 0, i_1, (i_2 i_1),\ldots, (i_n \ldots i_2 i_1),\ldots
\]
and the upper collective index
\[
       I= 0, i_1, (i_1 i_2),\ldots, (i_1 i_2\ldots i_n),\ldots
\]
are used in the last expression for compactness.
Therefore, the mapped vector can be rewritten as
\[
     x'= h(e_I)\cdot x^I\,.
\]
Introduce a decomposition of vectors $h(e_I)$ in terms of
the basis vectors $e_K$
\[
     h(e_I) =  e_K\cdot {h^K}\!_I\,.
\]
Using this relation we get
\begin{equation}
     x'=e_K\cdot {h^K}\!_I \cdot x^I \,.
\label{F01}
\end{equation}

Let $\B{U}$ be the set of linear mappings. We define an addition and a
multiplication by a number, satisfying the distributivity rule.  As a
result, $\B{U}$ becomes a vector space.

Introduce {\it basis mappings\/} ${\II^I}\!_K(\ )$ on the vector space $\B{U}$
so that
\[
     h(\ )={h^K}\!_I\cdot {\II^I}\!_K(\ ) \,,
\]
where {\it the mapping coordinates\/} ${h^K}\!_I$ are
the decomposition coefficients of the vectors $h(e_I)$ along
the basis vectors $e_K$. The linear mapping of the vector $x=e_L\cdot x^L$
should have the form \eref{F01}, therefore
\[
     {h^K}\!_I\cdot{\II^I}\!_K(e_L)\cdot x^L=
     e_K\cdot {h^K}\!_I\cdot x^I \,.
\]
From here, a mapping rule for the basis vectors $e_K$ of the space
$\B{X}$ through the basis mappings ${\II^I}\!_K(\ )$ follows
\begin{equation}
     {\II^I}\!_K(e_L)=e_K\cdot{\delta^I}\!_L\,,
\label{F02}
\end{equation}
where ${\delta^I}\!_L$ is the Kronecker delta.

\subsection{Linear mapping group. Linear mapping algebra}

We introduce {\it a group composition rule\/} acting on the space $\B{U}$:
\begin{equation}
     h(x)=h_2(h_1(x)) \,,\qquad\qquad  x\in\B{X}\,,
     \quad  h(\ ),h_1(\ ),h_2(\ )\in \B{U} \,,
\label{F03}
\end{equation}
i.e. we require that the set of linear mappings $\B{U}$ is {\it a group}.
Write the mappings involved in the composition rule through the basis mappings:
\[
     h(\ ) = {h^M}\!_L\cdot {\II^L}\!_M(\ )\,,\qquad
     h_2(\ ) = {(h_2)^M}\!_I\cdot {\II^I}\!_M(\ )\,,\qquad
     h_1(\ ) = {(h_1)^K}\!_L\cdot {\II^L}\!_K(\ )\,.
\]
From \eref{F03}, we obtain the relation between the coordinates
of these mappings:
\[
      {h^M}\!_L={(h_2)^M}\!_K\cdot{(h_1)^K}\!_L
\]
and the composition rule for the basis mappings
\[
      {\II^I}\!_M({\II^L}\!_K(\ )) = {\delta^I}\!_K\cdot{\II^L}\!_M(\ ) \,.
\]
The composition rule acting on vectors of the
space $\B{U}$ can be considered as a multiplication rule and can be written
in the algebraic form instead of the operational one:
\[
     h = h_1\circ h_2\,.
\]
The composition rule for the basis mappings is rewritten as
\begin{equation}
      {\II^L}\!_K\circ {\II^I}\!_M = {\delta^I}\!_K \cdot {\II^L}\!_M \,.
\label{F04}
\end{equation}
The linear mapping operation vector can also be considered as a
multiplication
\[
     x' = x\circ h\,.
\]
The composition rule \eref{F02} for basis vectors of $\B{X}$ and $\B{U}$
is rewritten as
\[
      e_L\circ {\II^I}\!_K = {\delta^I}\!_L \cdot e_K \,.
\]
We suppose that the composition and addition rules for linear mappings are
connected by the distributivity rule. Thereof the linear mapping vector space
$\B{U}$ is {\it an algebra}.

\subsection{Turn group of subspace of the generalized space-time}

Let $\B{D}$ be a subspace of the generalized space-time $\B{X}$.
If $\B{D}$ is an algebra, the composition rule for the basis
vectors $\varepsilon_I\in\B{D}$ has the form:
\[
      \varepsilon_I\circ \varepsilon_K = \varepsilon_L\cdot {\CC^L}\!_{IK} \,.
\]
Here ${\CC^L}\!_{IK}$ are the structural constants or
{\it the parastrophic matrices\/} of the algebra $\B{D}$.

Let us introduce {\it a scalar product\/} of vectors $x_1,x_2\in\B{D}$:
\[
     \langle x_1, x_2\rangle = \langle \varepsilon_I, \varepsilon_K \rangle
          \Cdot ({x_1})^I \Cdot ({x_2})^K =
     \g_{IK}\cdot ({x_1})^I\Cdot ({x_2})^K \,.
\]
The quantity $\g_{IK}=\varepsilon_0\cdot {\CC^0}\!_{IK}$ is
{\it the metric tensor}. Note that for the space of leptons $\B{C}_4$,
$\g_{IK}$ represents the diagonal matrix whose the diagonal is
the signature of basis vectors $\varepsilon_I$.
The scalar product of vector by itself is the vector {\it length\/}:
\[
     \langle x,x\rangle= \g_{IK}\cdot x^I \cdot x^K = x^2 \,.
\]

Consider vectors $x'_1,x'_2\in\B{D}$ resulted from a linear mapping $h$ of
vectors $x_1,x_2\in\B{D}$. The linear mapping
changes the scalar product of vectors in a common case.
We extract from all linear mappings {\it rotations\/} which preserve the
scalar product in $\B{D}$:
\begin{equation}
     \langle {x_2}',{x_1}'\rangle = \langle {x_2},{x_1}\rangle\,.
\label{F05}
\end{equation}
From the condition \eref{F05} it follows that
the linear mapping matrix ${h^L}\!_I$ for rotations should satisfy
\[
     \g_{LM}\cdot {h^L}\!_I \cdot {h^M}\!_K\cdot \g^{KN} = {\delta_I}\!^N \,.
\]
If we introduce {\it a conjugate matrix\/} ${{\tld{h}}^N}\!_L$ as
\[
     {{\tld{h}}^N}\!_L = \g_{LM}\cdot {h^M}\!_K \cdot \g^{KN} \,,
\]
the condition that the linear mapping is rotation takes the form
\[
     {{\tld{h}}^N}\!_L  = (h^{-1}){}^N{}\!_L \,.
\]

\subsection{Parametrical representation of linear mappings}
\label{Turn_angles}

Consider vectors $h\in \B{U}$ as functions of {\it parameters\/}
$\varphi^\alpha$:
\[
      h(\varphi^\alpha) = {\II^I}\!_K \cdot {h^K}\!_I (\varphi^\alpha) \,.
\]
We suppose that the group composition rule acts on parameters
$\varphi^\alpha$:
\[
      \varphi^\alpha= \Phi({\varphi_2}^\alpha, {\varphi_1}^\alpha) \,,
\]
and the correspondence exists between the composition rule on
$\{\varphi^\alpha\}$ and the multiplication rule on $\B{U}$:
\[
      h(\varphi^\alpha) = h({\varphi_1}^\alpha) \circ h({\varphi_2}^\alpha) \,,
\]
and the units of both groups are also in correspondence to one another
\[
      \left. {h^K}\!_I (\varphi^\alpha) \right|_{\varphi^\alpha=0}
      = {\delta^K}\!_I \,.
\]
For the turn group, such parameters are called {\it turn angles}.
Consider a differential $\dd h$ for $h$ close to the group unit:
\[
       \dd h(\varphi^\alpha) = {\II^I}\!_K \Cdot
       \DD{{h^K}\!_I (\varphi^\alpha)}{\varphi^\alpha} \Cdot
       \dd \varphi^\alpha =
       {\II^I}\!_K \cdot {\KK^K}\!_{I\alpha} \cdot \dd \varphi^\alpha \,.
\]
Here the notation
\[
       {\KK^K}\!_{I\alpha} = \left.
       \DD{{h^K}\!_I (\varphi^\alpha)}{\varphi^\alpha}
       \right|_{\varphi^\alpha=0}
\]
was introduced. The vectors
\begin{equation}
       \II_\alpha = {\II^I}\!_K\cdot {\KK^K}\!_{I\alpha}
\label{F06}
\end{equation}
are the basis vectors in the space of vectors of the type
$\dd h = \II_\alpha \Cdot \dd \varphi^\alpha$.
The multiplication rule for these basis vectors has the form
\begin{equation}
       \II_\alpha \circ \II_\beta =
       \II_\gamma\cdot {\CC^\gamma}\!_{\alpha\beta} \,.
\label{F07}
\end{equation}
If we substitute \eref{F06} in \eref{F07}
and take into account \eref{F04} we obtain
\[
       {\KK^K}\!_{L\beta} \cdot {\KK^L}\!_{I\alpha} =
       {\KK^K}\!_{I\gamma}\cdot {\CC^\gamma}\!_{\alpha\beta}  \,.
\]
Comparing this relation with \eref{F07}, we conclude that the basis
vectors $\II_\alpha$ can be put into the correspondence with
the matrices ${\KK^L}\!_{I\alpha}$:
\[
     \II_\alpha \sim {\KK^L}\!_{I\alpha}\,.
\]
This correspondence will be called {\it a parametrical representation\/}
of the linear mapping algebra.

\subsection{Turns in the Clifford algebra}

For the Clifford algebra, the turn matrix around the axis passing
through the origin and parallel to the vector $\varepsilon_\alpha$,
can be written as
\[
      {h^K}\!_I (\varphi^\alpha) =
      \left\{ \begin{array}{ll}
      {\delta^K}\!_I \cdot \cos \varphi^\alpha  +
      {\CC^K}\!_{I\alpha} \cdot \sin \varphi^\alpha\,,\qquad
      &\hbox{for }(\varepsilon_\alpha)^2=-1 \,; \\
      {\delta^K}\!_I \cdot \cosh \varphi^\alpha  +
      {\CC^K}\!_{I\alpha} \cdot \sinh \varphi^\alpha\,,
      &\hbox{for }(\varepsilon_\alpha)^2=1 \,, \\
      \end{array}\right.
\]
where ${\CC^K}\!_{I\alpha}$ are the parastrophic matrices of the regular
representation of basis vectors $\varepsilon_\alpha$ (see \cite{art1}),
there is no summation over the index $\alpha$.  From here
it follows that
\begin{equation}
       {\KK^K}\!_{I\alpha} = \left.
       \DD{{h^K}\!_I (\varphi^\alpha)}{\varphi^\alpha}
       \right|_{\varphi^\alpha=0} = {\CC^K}\!_{I\alpha}
\label{F08}
\end{equation}
for turns in the Clifford algebra.
If we use an {\it inverse\/} regular representation, i.e. the correspondence
of basis vectors to parastrophic matrices, we obtain the representation of
turns in the Clifford algebra through the basis vectors:
\[
      h(\varphi^\alpha) =
      \left\{ \begin{array}{ll}
      \varepsilon_0\Cdot \cos \varphi^\alpha  +
      \varepsilon_\alpha \Cdot \sin \varphi^\alpha\,,\qquad
      &\hbox{for }(\varepsilon_\alpha)^2=-1 \,; \\
      \varepsilon_0\Cdot \cosh \varphi^\alpha  +
      \varepsilon_\alpha \Cdot \sinh \varphi^\alpha\,,
      &\hbox{for }(\varepsilon_\alpha)^2=1 \,. \\
      \end{array}\right.
\]
Thus
\[
     \II_\alpha = {\II^I}\!_K \cdot {\CC^K}\!_{I\alpha} = \varepsilon_\alpha\,.
\]

\subsection{Gauge group. Gauge field}

Let us suppose that vectors $h\in\B{U}$ are functions of vectors $x\in\B{X}$.
In this case, the linear mappings will be called {\it gauge transformations}.
We shall assume that the gauge transformation group is {\it responsible for
interaction}. The function
\[
     h(x)={\II^I}\!_K\cdot{h^K}\!_I(x)
\]
will be named a {\it gauge $h$-field}.

A differential of transformation $h''(x)= h'(x)\circ h(x)$ is
\[
     \dd h'' = \dd h'\circ h + h'\circ \dd h \,.
\]
We multiply this expression on the inverse vector
$h''^{-1}=h^{-1}\circ h'^{-1}$ at the left
\begin{equation}
     h''^{-1}\circ \dd h'' = h^{-1}\circ h'^{-1}\circ \dd h'\circ h +
     h^{-1}\circ \dd h \,.
\label{F09}
\end{equation}
Introduce a function
\[
     {A^K}\!_{IM}(x) = (h^{-1}){}^K{}\!_L\Cdot \DD{h{}^L{}\!_I}{x^M} \,,
\]
which will be called {\it a gauge field  potential\/} for an arbitrary gauge
transformation. From \eref{F09} we obtain a common transformation rule for
the potential
\[
     {A''^K}\!_{IM} =
     (h^{-1}){}^K{}\!_N\cdot {A'^N}\!_{LM}\cdot {h}{}^L{}\!_I
     + {A^K}\!_{IM} \,.
\]
The above definition of the potential and its transformation rule
are simplified when the gauge transformations are close
to the unit transformation. In this case $\dd h''=\dd h'+\dd h$,
\[
        {A^K}\!_{IM} = \DD{h{}^K{}\!_I}{x^M}   \,,\qquad\qquad
        {A''^K}\!_{IM} = {A'^K}\!_{IM} + \DD{{h^K}\!_I}{x^M} \,,
\]
and in the parametrical representation
\[
        {A^K}\!_{IM} =
       \left.
       \DD{{h^K}\!_I (\varphi^\alpha)}{\varphi^\alpha}
       \right|_{\varphi^\alpha=0} \Cdot \DD{\varphi^\alpha}{x^M} =
       {\KK^K}\!_{I\alpha}\cdot {A^\alpha}\!_{M} \,.
\]

\subsection{Structure equations of the generalized space-time in a gauge
field}

We shall further restrict our consideration to particles, but in the
conclusions we shall discuss how antiparticles can be described together with
particles.  Introduce a vector space $\B{T}=\B{X}+\B{U}$ which will be called
{\it kinematic}.  Besides the multiplications $\B{X}\circ \B{X} \to \B{X}$,
$\B{U}\circ \B{U} \to \B{U}$ and $\B{X}\circ \B{U} \to \B{X}$, we define the
multiplication $\B{U}\circ \B{X} \to 0$.  Thus we shall use the following
multiplication rules for the basis vectors:
\begin{mathletters}
\begin{eqnarray}
\label{F_rules}
     \varepsilon_I \circ \varepsilon_K
          &=& \varepsilon_L \cdot {\CC^L}\!_{IK} \,,\\
     {\II^L}\!_K \circ {\II^I}\!_M
          &=& {\II^L}\!_M \cdot {\delta^I}\!_K   \,,\\
     \varepsilon_L \circ {\II^I}\!_K
          &=& \varepsilon_K \cdot {\delta^I}\!_L \,,\\
     {\II^I}\!_M \circ \varepsilon_K &=& 0       \,.
\label{F_last_rule}
\end{eqnarray}
\end{mathletters}
As a result, the kinematic space $\B{T}$ becomes algebra. Note that the
simplest rule \eref{F_last_rule} is necessary for closing the kinematic
algebra.
We write the multiplication rule for vectors $t, t_1, t_2 \in \B{T}$ as
\begin{equation}
      t= t_1\circ t_2\,.
\label{F10}
\end{equation}

For algebras, there are typical differential relations resulting from
the derivation of the multiplication rule. These relations are
called {\it the structure equations}. It was shown in our previous paper
\cite{art1} that the quantum mechanics equations for free particles
can be derived from the structure equations for subalgebras of
the generalized space-time $\B{X}$ and those of the action space $\B{SX}$.
We shall apply this approach to the kinematic algebra $\B{T}$.

Differential of a vector $t$ with variation in a vector $t_i$ will be denoted
by $\D_i t$.  The double derivation of the multiplication rule \eref{F10}, at
first by $t_1$ and next by $t_2$, gives the common structure equation of the
kinematic algebra:
\[
     \D_2 \D_1 t = \D_1 t \circ (t)^{-1} \circ \D_2 t \,.
\]
For $t$ close to the group unit, it is reduced to
\[
      \D_2 (\D_1 x + \D_1 h) = (\D_1 x + \D_1 h)\circ (\D_2 x + \D_2 h) \,,
\]
where it was taken into account that $t=x+h$.
Let us consider the {\it projection\/} of the last relation on the generalized
space-time:
\[
      \D_2 \D_1 x = \D_1 x \circ \D_2 t \,.
\]
This equation will be called {\it the structure equation of the
generalized space-time in a gauge field}.

\section{Relativistic quantum mechanics equations for particles
in a gauge field}

\subsection{Action space and its linear mappings}

In our book \cite{book}, the ability of bodies to interact was associated
with the presence of {\it the action vector\/} $S$.  Such vectors form the
vector space $\B{SX}$. We have assumed that the space $\B{SX}$ is {\it
similar} to the space $\B{X}$, bearing in mind that basis vectors of the
space $\B{X}$ can be used as basis vectors in the space $\B{SX}$. Therefore
the vector $S\in \B{SX}$ can be written as $S=e_N\cdot S^N$.  $\B{SX}$ is
algebra as well as $\B{X}$ with the same multiplication rule for basis
vectors.

In the present work we expand the notion of the action vector $S$.
Let us consider that the action space $\B{S}$ is the sum
\[
     \B{S} = \B{SX} + \B{SU}\,,
\]
and the vector spaces $\B{SX}$ and $\B{SU}$ are similar to the spaces
$\B{X}$ and $\B{U}$, respectively.
In other words, the action vector $S\in \B{S}$ is represented by the sum
of two component
\[
     S = S_x + S_h \,.
\]
Here $S_x=e_N\cdot S^N\in \B{SX}$, and
$S_h={\II^L}\!_I\cdot {S^I}\!_L\in \B{SU}$.

We consider the action vector as a function of generalized space-time vector
$x$ and gauge $h$-field: $S=S(x,h)$. Let us suppose that the
coordinates $S^K$ depend only on $x$, and the coordinates ${S^K}\!_L$ depend
only on $h(\varphi^\alpha)$, that is
\[
     S(x,\varphi^\alpha) = e_N\cdot S^N(x) +
     {\II^L}\!_I\cdot {S^I}\!_L(\varphi^\alpha) \,.
\]
Consider a differential of action vector
\[
     \dd S = \frac{\partial S}{\partial x^M} \Cdot \dd x^M +
     \frac{\partial S}{\partial \varphi^\alpha} \Cdot \dd \varphi^\alpha \,.
\]
Let us introduce {\it a generalized impulse}
\[
     p_M \equiv - \frac{\partial S}{\partial x^M} =
     - \frac{\partial S_x}{\partial x^M} =
     - e_N\Cdot \frac{\partial S^N}{\partial x^M} = e_N\cdot {p^N}\!_M  \,,
\]
and {\it a generalized moment}
\[
     m_\alpha \equiv - \frac{\partial S}{\partial \varphi^\alpha} =
     - \frac{\partial S_h}{\partial \varphi^\alpha} =
     - {\II^L}\!_I\Cdot \frac{\partial {S^I}\!_L}{\partial \varphi^\alpha} =
     {\II^L}\!_I\cdot {m^I}\!_{L\alpha}  \,.
\]
Thus,
\[
     \dd S = - p_M \cdot \dd x^M - m_\alpha \cdot \dd \varphi^\alpha \,.
\]

We discuss the parametrical representation of vectors $S_h\in \B{SU}$.
In line with the Section \ref{Turn_angles}, we suppose that the vectors
$S_h$ are functions of parameters $\SC^\alpha$ with dimensionality of
action:
\[
      S_h(\SC^\alpha) = {\II^L}\!_I \cdot {S^I}\!_L (\SC^\alpha) \,,
\]
and the group composition rule acts on the parameters $\SC^\alpha$
similar to the parameters $\varphi^\alpha$:
\[
      \SC^\alpha= \Phi({\SC_2}^\alpha, {\SC_1}^\alpha) \,.
\]
We also assume that the correspondence exists between the composition rule on
$\{\SC^\alpha\}$ and the multiplication rule on $\B{SU}$:
\[
      S_h (\SC^\alpha) =
      S_h ({\SC_1}^\alpha) \circ S_h ({\SC_2}^\alpha) \,,
\]
and the units of both groups are in correspondence to one another
\[
      \left. {S^I}\!_L (\SC^\alpha) \right|_{\SC^\alpha=0}
      = S^0 \Cdot {\delta^I}\!_L \,,
\]
where the scalar component of action vector $S^0$ is action
in a classical sense. Let us consider a differential
\[
       \dd S_h(\SC^\alpha) = {\II^L}\!_I \Cdot
       \DD{{S^I}\!_L (\SC^\alpha)}{\SC^\alpha} \Cdot \dd \SC^\alpha
\]
for $S_h$ close to the group unit.
From the similarity of the spaces $\B{SU}$ and $\B{U}$ it follows that
\[
       \left.
       \DD{{S^I}\!_L (\SC^\alpha)}{\SC^\alpha}
       \right|_{\SC^\alpha=0} = {\KK^I}\!_{L\alpha} \,.
\]
Then
\[
       \dd S_h (\SC^\alpha) =
       {\II^L}\!_I \cdot {\KK^I}\!_{L\alpha} \cdot \dd \SC^\alpha =
       \II_\alpha \Cdot\dd \SC^\alpha \,,
\]
where the relation \eref{F06} was used. The vector $\II_\alpha$ can be
considered as the basis one in the space of vectors of the type
$\dd S_h = \II_\alpha \cdot \dd \SC^\alpha$.
In the parametrical representation, the gauge
$h$-field can be expressed through angles $\varphi^\alpha$. The coordinates
${S^I}\!_L(h)$ can also be written as functions of angles $\varphi^\alpha$:
${S^I}\!_L={S^I}\!_L(h(\varphi^\alpha))$. Therefore, one can say
about the parametrical representation of action vector $S_h$.
The generalized moment coordinates are written in the parametrical
representation as
\begin{equation}
     {m^I}\!_{L\beta} = \DD{{S^I}\!_L(h)}{\varphi^\beta} =
     {\KK^I}\!_{L\alpha}\Cdot \DD{\SC^\alpha (\varphi)}{\varphi^\beta} =
     {\KK^I}\!_{L\alpha}\cdot {g^\alpha}\!_\beta \,.
\label{F11}
\end{equation}
Functions ${g^\alpha}\!_\beta\equiv \partial\SC^\alpha/\partial\varphi^\beta$
form {\it a coupling matrix}.

We assume that any {\it type of interactions\/} can be associated with
some {\it subgroup\/} of the gauge group. Let us consider that the
subgroup of $i$-th type of interactions has a single coupling coefficient
$g_i$ which will be called {\it a gauge charge\/} of this type of
interactions. In other words, we suppose that the
following relation is fulfilled
\begin{equation}
     {g^{\alpha_i}}\!_{\beta_i} = g_i\cdot {\delta^{\alpha_i}}\!_{\beta_i} \,.
\label{F12}
\end{equation}
Thus the gauge charge has a meaning of the coefficient of similarity between
the parameters $\SC^\alpha$ and $\varphi^\alpha$ used for
the representation of vectors $S_h$ and $h$.
Using \eref{F11} and \eref{F12} we get
\begin{equation}
     \DD{{S^I}\!_L}{x^M} =
     {\KK^I}\!_{L\alpha} \Cdot \DD{\SC^\alpha}{x^M} =
     {\KK^I}\!_{L\alpha} \Cdot \DD{\SC^\alpha}{\varphi^\beta}\Cdot
     \DD{\varphi^\beta(x)}{x^M}
     = {\KK^I}\!_{L\alpha} \cdot {g^\alpha}\!_\beta\cdot {A^\beta}\!_M =
     \sum_i g_i\Cdot {\KK^I}\!_{L\alpha_i} \cdot {A^{\alpha_i}}\!_M \,,
\label{F13}
\end{equation}
where ${\KK^I}\!_{L\alpha_i}$ are the parastrophic matrices of $i$-th
subgroup of gauge transformations; ${A^{\alpha_i}}\!_M$ is the gauge field
potential appropriate to this subgroup; the summation is over all
interactions.
From this relation, we obtain the potential expressed through
the parameter coordinates $\SC^\alpha$:
\[
        {A^I}\!_{LM} = \frac{1}{g} \Cdot \DD{{S^I}\!_L}{x^M} =
        \frac{1}{g} \Cdot {\KK^I}\!_{L\alpha} \Cdot \DD{\SC^\alpha}{x^M} \,,
\]
and the transformation rule for the potential:
\[
        {A''^I}\!_{LM} = {A'^I}\!_{LM}
        + \frac{1}{g} \Cdot \DD{{S^I}\!_L}{x^M}\,.
\]

As well as the algebra $\B{T}$, the algebra $\B{S}$ has the structure equation
\[
      \D_2 \D_1 S = - \frac{1}{S^0}\Cdot \D_1 S\circ \D_2 S\,.
\]
After the projecting on the action subspace $\B{SX}$, this equation
takes the form:
\begin{equation}
      \D_2 \D_1 S_x = - \frac{1}{S^0}\Cdot \D_1 S_x \circ \D_2 S \,.
\label{F14}
\end{equation}
It will be called {\it the structure equation of the action space $\B{SX}$
in a gauge field}.  Let us also write the structure equations in
the coordinate form
\begin{equation}
      \D_2 \D_1 S^I  =- \frac{1}{S^0}\Cdot
      ({\CC^I}\!_{LN}\cdot \D_2 S^N \cdot \D_1 S^L  +
      \D_2 {S^I}\!_L\cdot \D_1 S^L)\,.
\label{F15}
\end{equation}

\subsection{Quantization equations in differentials}

Hereafter, we shall use the system of natural units ($\hbar=c=1$).

We perform the passage to the dynamic equations of quantum mechanics
by according to the new interpretation of {\it the wave function}.
In \cite{book,art1}, the wave function was interpreted as
a differential of action vector
\[
     \psi=\D_1 S_x \,.
\]
In the equations \eref{F14} and \eref{F15}
we denote the differential $\D_2$ by $\dd$. Let us
set $S^0=\hbar=1$, i.e. we shall consider the equations for
action vector values close to the Planck constant.  The structure
equations with respect to the wave function
\[
      \dd \psi + \psi\circ \dd S_h = - \psi\circ \dd S_x
\]
or with respect to its coordinates
\begin{equation}
      \dd \psi^I +  \dd {S^I}\!_L\cdot \psi^L
      = - {\CC^I}\!_{LN}\cdot \dd S^N \cdot \psi^L
\label{F16}
\end{equation}
will be called {\it quantization equations in differentials}.

We shall suppose that action vectors and their linear
transformations are functions of generalized space-time vectors.
Then from the quantization equations in differentials \eref{F16}, the
relations follow
\[
     \partial_M \psi^I(x) + \partial_M{S^I}\!_L\cdot \psi^L
      = {\CC^I}\!_{LN}\cdot {p^N}\!_M  \cdot \psi^L \,,
\]
where ${p^L}\!_M=-\partial_M S^L$ are the generalized impulse coordinates.
These equations will be named {\it quantum postulates for particles in a gauge
field}.
Using the relations \eref{F13} the quantum postulates take the form
\begin{equation}
      \partial_M\psi^I(x) +
      {\KK^I}\!_{L\alpha}\cdot {g^\alpha}\!_\beta\cdot
      {A^\beta}\!_M\cdot\psi^L =
      {\CC^I}\!_{LN}\cdot {p^N}\!_M \cdot \psi^L \,.
\label{F17}
\end{equation}

\subsection{Relativistic quantum mechanics equations for particles
in a gauge field}

Multiply the quantum postulates \eref{F17} by ${\CC^{MK}}\!_I$:
\[
     {\CC^{MK}}\!_I \Cdot \left(\partial_M\psi^I(x) +
     {\KK^I}\!_{L\alpha}\cdot {g^\alpha}\!_\beta\cdot
     {A^\beta}\!_M\cdot\psi^L \right) =
     {\CC^{MK}}\!_I\cdot
     {\CC^I}\!_{LN} \cdot {p^N}\!_M  \cdot \psi^L\,.
\]
These equations will be called {\it relativistic quantum mechanics equations
in Dirac's form for particles in a gauge field}.  We suppose that the wave
function $\psi(x)$ depends only on coordinates of the space-time $X$.  We
pass from the generalized action space $\B{S}$ to the space of leptons
$\B{SC}_4$ and to the turn group in this space. Then we obtain
\begin{equation}
     {\CC^{mK}}\!_I \cdot \partial_m\psi^I(x) +
     {\CC^{MK}}\!_I \cdot {\CC^I}\!_{L\alpha}\cdot
     {g^\alpha}\!_\beta\cdot {A^\beta}\!_M\cdot\psi^L =
     {\CC^{MK}}\!_I\cdot
     {\CC^I}\!_{LN} \cdot {p^N}\!_M  \cdot \psi^L\,,
     \qquad (m=1,\ldots,4) \,.
\label{F18}
\end{equation}
Here $\psi^I(x)$ are sixteen real functions, ${\CC^{mK}}\!_I$ are the regular
representation matrices of basis vectors of the space-time $X$ in the space
of leptons $\B{C}_4$ and, in addition, the relation \eref{F08} was used.
These equations will be named the relativistic quantum mechanics equations
{\it for the leptons\/} in a gauge field.

According to \cite{art1}, we assume that the
generalized impulse ${p^N}\!_M$ has only the two components
\[
     {p^0}\!_0     {}= - \partial_0 S^0    = \frac{\mc}{2}  \,,\qquad\qquad
     {p^{34}}\!_0  {}= - \partial_0 S^{34} = \frac{\mc}{2}  \,.
\]
If we substitute these components in \eref{F18} and use the
relations $C^{0K}\!{}_I=\delta^K\!{}_I$, $C^I\!{}_{L0}=\delta^I\!{}_L$, and
\eref{F12}, we obtain the quantum mechanics equations for the
leptons in a gauge field in the final form:
\begin{equation}
     {\CC^{mK}}\!_I \cdot \partial_m\psi^I(x) +
     {\CC^{MK}}\!_I \Cdot \sum_i g_i \Cdot {\CC^I}\!_{L\alpha_i}\cdot
     {A^{\alpha_i}}\!_M\cdot\psi^L =
     \frac{\mc}{2}\Cdot
     ({\delta^K}\!_L + {\CC^K}\!_{L34}) \Cdot \psi^L\,.
\label{F19}
\end{equation}

\subsection{Quantum mechanics equations for the leptons in the electromagnetic
field}

We assume that the direct product of the turn groups
about the axes $\varepsilon_{21}$ and $\varepsilon_{1324}$:
\[
     h_1 = \varepsilon_0\Cdot\cos \varphi_{21} +
     \varepsilon_{21}\Cdot\sin \varphi_{21} \,,\qquad\qquad
     h_2 = \varepsilon_0\Cdot\cos \varphi_{1324} +
     \varepsilon_{1324}\Cdot\sin \varphi_{1324} \,,
\]
in the space of leptons $\B{C}_4$
is responsible for the {\it electromagnetic\/} interaction of leptons.
Each of the groups is isomorphic to the turn group $U(1)$ in the complex plane.
Therefore the group, responsible for the electromagnetic interaction of
leptons, is isomorphic to the product $U(1)\times U(1)$. This group
will be called {\it electrical}.  Thus, in order to describe the leptons
in the electromagnetic field we set in the equations \eref{F19}
\[
     \alpha_i=21,1324\,, \qquad\qquad g_i \equiv\frac{e}{2}\,,
\]
where $e$ is the elementary electric charge.
Let us assume that the angles $\varphi_{21}$ ¨ $\varphi_{1324}$ depend
only on coordinates of the space-time $X$. In other words, the
electromagnetic field is described by the potential components
\[
     {A^{\alpha_i}}\!_M = \DD{\varphi^{\alpha_i}}{x^M} =
     \{{A^{21}}\!_m,\,{A^{1324}}\!_m\}\,.
\]
In the space of the electrical group, we select the section
$\varphi_{1324}= \varphi_{21}\equiv\varphi$. Then
\[
     {A^{21}}\!_m = {A^{1324}}\!_m \equiv \AAA_m \,,
\]
where $\AAA_m$ are the electromagnetic potential components.
As a result, the equations \eref{F19} are reduced to
the quantum mechanics equations for the leptons in the electromagnetic field:
\[
     {\CC^{mK}}\!_I \Cdot \left(\partial_m \psi^I +
     \frac{e\Cdot \AAA_m}{2}\Cdot
     ({\CC^I}\!_{L {21}} + {\CC^I}\!_{L {1324}})\Cdot \psi^L\right)
      {}=
     \frac{\mc}{2}\Cdot ({\delta^K}\!_L + {\CC^K}\!_{L34})
      \Cdot \psi^L\,.
\]
Let us now pass to the quaternion representation of the parastrophic matrices
and that of the wave functions (see \cite{art1}) for which
{\renewcommand{\arraystretch}{0.95}
\tighten
\tabcolsep = 1mm
\[
      {\CC^I}\!_{L {21}} = i \,\Cdot
      \hbox{\footnotesize
      \begin{tabular}{|cc|cc|}
        \hline
        \p 1 &     &     &     \\
             &\p 1 &     &     \\
        \hline
             &     &\p 1 &     \\
             &     &     &\p 1 \\
        \hline
      \end{tabular}}
      \,,\qquad {\CC^I}\!_{L {1324}} = i \,\Cdot
      \hbox{\footnotesize
      \begin{tabular}{|cc|cc|}
        \hline
              &\p 1 &     &     \\
        \p 1  &     &     &     \\
        \hline
              &     &     &\p 1 \\
              &     &\p 1 &     \\
        \hline
      \end{tabular}} \,.
\]
}%
For the leptons of the first generation in the electromagnetic field, we obtain
{\tighten
\[
     ||{\CC^{mK}}\!_I|| \Cdot \left( \partial_m \Cdot
     \hbox{\footnotesize
     $\begin{array}{||l||}
     \Psi^{\scriptscriptstyle 0  }   \\
     \Psi^{\scriptscriptstyle 34 }   \\
     \Psi^{\scriptscriptstyle 123}   \\
     \Psi^{\scriptscriptstyle 124}   \\
     \end{array}$}
     \, + \, \frac{i\Cdot e\Cdot \AAA_m}{2}\Cdot
     \hbox{\footnotesize
     $\begin{array}{||l||}
     \Psi^{\scriptscriptstyle 0  }+\Psi^{\scriptscriptstyle 34 }   \\
     \Psi^{\scriptscriptstyle 0  }+\Psi^{\scriptscriptstyle 34 }   \\
     \Psi^{\scriptscriptstyle 123}+\Psi^{\scriptscriptstyle 124}   \\
     \Psi^{\scriptscriptstyle 123}+\Psi^{\scriptscriptstyle 124}   \\
     \end{array}$}
     \right)
     = \frac{\mc}{2} \, \Cdot
     \hbox{\footnotesize
     $\begin{array}{||l||}
     \Psi^{\scriptscriptstyle 0  }+\Psi^{\scriptscriptstyle 34 }   \\
     \Psi^{\scriptscriptstyle 0  }+\Psi^{\scriptscriptstyle 34 }   \\
     \Psi^{\scriptscriptstyle 123}+\Psi^{\scriptscriptstyle 124}   \\
     \Psi^{\scriptscriptstyle 123}+\Psi^{\scriptscriptstyle 124}   \\
     \end{array}$} \;.
\]
}%
Here the matrices
${\CC^{mK}}\!_I = \{{\CC^{aK}}\!_I\,,{\CC^{4K}}\!_I\}$ ($a=1,2,3$)
have the form:
{\renewcommand{\arraystretch}{0.95}
\tighten
\tabcolsep = 1mm
\[
      {\CC^{aK}}\!_I = i\,\Cdot
      \hbox{\footnotesize
      \begin{tabular}{|cc|cc|}
        \hline
                      &              &\mL{\sigma^a} &                 \\
                      &              &              &\mL{\sigma^a}    \\
        \hline
        \pL{\sigma^a} &              &              &                 \\
                      &\pL{\sigma^a} &              &                 \\
        \hline
      \end{tabular}}
      \,,\qquad
      {\CC^{4K}}\!_I = i \,\Cdot
      \hbox{\footnotesize
      \begin{tabular}{|cc|cc|}
        \hline
            &    &    &\p 1\\
            &    &\p 1&    \\
        \hline
            &\p 1&    &     \\
        \p 1&    &    &     \\
        \hline
      \end{tabular}} \,.
\]
}%
In order to obtain the equations with respect to the right and left
components of electron and those of $e$-neutrino, we transform these
equations by the same way as was made in \cite{art1}. At first we add the
first equation with the second one and the third one with the fourth one.
Then we subtract the third equation from the fourth one, and the first one
from the second one.  Taking into account that
\begin{eqnarray*}
     \Psi^0 + \Psi^{34} = e\LL&&
     \hbox {, the left component of electron,} \\
     \Psi^{123} + \Psi^{124} = e\RR&&
     \hbox {, the right component of electron,} \\
     \Psi^{123} - \Psi^{124} = \nu_e\LL&&
     \hbox {, the left component of $e$-neutrino,} \\
     \Psi^0 - \Psi^{34} = \nu_e\RR&&
     \hbox {, the right component of $e$-neutrino,}
\end{eqnarray*}
we obtain the quantum mechanics equations for the leptons of the first
generation in the electromagnetic field
\begin{eqnarray*}
    i\Cdot \gamma_1^m \left(
    \partial_m  + i\Cdot e\Cdot \AAA_m
    \right) e\RR &=& \mc \Cdot e\LL \,, \\
    i\Cdot \gamma_2^m \left(
    \partial_m  + i\Cdot e\Cdot \AAA_m
    \right) e\LL &=& \mc \Cdot e\RR \,, \\
     i\Cdot \gamma_1^m \Cdot \partial_m \nu_e\RR &=& 0 \,, \\
     i\Cdot \gamma_2^m \Cdot \partial_m \nu_e\LL &=& 0 \,.
\end{eqnarray*}
Here
\[
     \gamma_1^m = \{-\sigma^a \,, 1\}\,, \qquad\qquad
     \gamma_2^m = \{ \sigma^a \,, 1\}\,.
\]
Thus, the system of four equations is transformed to the two independent
systems of two equations.
As one would expect, the right and left electrons interact with the
electromagnetic field with the identical coupling constant, and the neutrino
does not interact with the electromagnetic field.

\subsection{Quantum mechanics equations for the
leptons in the electroweak field}

\subsubsection{The first approximation}

The interaction of leptons with the electromagnetic field
is considered by according to the previous Section.

We assume that the direct product of the turn groups about the axes
$\varepsilon_{1324}$, $\varepsilon_4$, $\varepsilon_{123}$,
$\varepsilon_{34}$, $\varepsilon_{124}$, $\varepsilon_{3}$:
\[
\begin{array}{l@{\qquad\qquad}l}
     h_1 = \varepsilon_0\Cdot\cos \varphi_{1324} +
     \varepsilon_{1324}\Cdot\sin \varphi_{1324} \,,&
     h_4 = \varepsilon_0\Cdot\cosh \varphi_{34} +
     \varepsilon_{34}\Cdot\sinh\varphi_{34}   \,, \\
     h_2 = \varepsilon_0\Cdot\cos \varphi_4 +
     \varepsilon_4\Cdot\sin \varphi_4           \,,&
     h_5 = \varepsilon_0\Cdot\cosh\varphi_{124} +
     \varepsilon_{124}\Cdot\sinh\varphi_{124} \,, \\
     h_3 = \varepsilon_0\Cdot\cos \varphi_{123} +
     \varepsilon_{123}\Cdot\sin \varphi_{123}   \,,&
     h_6 = \varepsilon_0\Cdot\cosh\varphi_3 +
     \varepsilon_3\Cdot\sinh\varphi_3
\end{array}
\]
in the space of leptons $\B{C}_4$ is responsible for the {\it weak\/}
interaction of leptons.  This group will be called {\it weak}. In this
approximation, the turn group about the axis $\varepsilon_{1324}$ is
responsible for the mixed electroweak interaction with $\AAA$ and $Z$
fields.  Note that the weak group is isomorphic to the Lorentz group
and can be represented as the product $SU(2)\times SU(2)$.

Thus, in order to describe the interaction of leptons with the weak field,
we should set in \eref{F19}
\[
     \alpha_i=1324,4,123,34,124,3 \,, \qquad\qquad g_i = \frac{g_\W}{2} \,.
\]
The gauge charge of the weak group is given by the constant $g_\W$.
We suppose that the angles $\varphi_{1324}$, $\varphi_{4}$, $\varphi_{123}$
depend on the space-time coordinates, and the angles $\varphi_{34}$,
$\varphi_{124}$, $\varphi_{3}$ depend on the coordinates $x^{234}$,
$x^{134}$, $x^{124}$, $x^{123}$ which can be written as $x^{m1324}$.  From
here follows that the weak field is described by the potential components
\[
     {A^{\alpha_i}}\!_M = \DD{\varphi^{\alpha_i}}{x^M} =
     \{{A^{1324}}\!_m,\,{A^{4}}\!_m,\,{A^{123}}\!_m,\,
       {A^{34}}\!_{m1324},\,{A^{124}}\!_{m1324},\,{A^{3}\!_{m1324}}\}\,.
\]
We postulate the following correspondences:
\[
     g_{1324}\Cdot {A^{1324}}\!_{m} \equiv
          \frac{e}{2}\Cdot \AAA_m + \frac{g_\W}{2} \Cdot Z_m \,, \qquad
     {A^{34}}\!_{m1324} \equiv Z_m  \,, \qquad
     {A^{4}}\!_{m} = {A^{124}}\!_{m1324}   \equiv W_m^1\,, \qquad
     {A^{123}}\!_{m} = {A^{3}}\!_{m1324}   \equiv W_m^2\,,
\]
where $Z_m$, $W_m^1$, $W_m^2$ are the weak field potentials.

As a result, the equations \eref{F19} is reduced to
the quantum mechanics equations for the leptons in the electroweak field:
\begin{eqnarray*}
     &&{\CC^{mK}}\!_I \Cdot \left( \partial_m \psi^I +
     \frac{e\Cdot \AAA_m}{2}\Cdot
     ({\CC^I}\!_{L{21}} + {\CC^I}\!_{L{1324}})\Cdot \psi^L +
     \frac{g_\W}{2}\Cdot
     (Z_m\Cdot {\CC^I}\!_{L{1324}} + W^1_m\Cdot {\CC^I}\!_{L{4}} +
     W^2_m\Cdot {\CC^I}\!_{L{123}})\Cdot \psi^L \right) \\
     &&{}+
     {\CC^{1324mK}}\!_P \Cdot \frac{g_\W}{2}\Cdot
     (Z_m\Cdot {\CC^P}\!_{L{34}} + W^1_m\Cdot {\CC^P}\!_{L{124}} +
     W^2_m\Cdot {\CC^P}\!_{L{3}})\Cdot \psi^L
     = \frac{\mc}{2}\Cdot
     ({\delta^K}\!_L + {\CC^K}\!_{L34}) \Cdot \psi^L\,.
\end{eqnarray*}
If we write the matrix ${\CC^{1324mK}}\!_P$ as the product
\[
     {\CC^{1324K}}\!_I\cdot {\CC^{mI}}\!_P =
     - {\CC^{mK}}\!_I\cdot {\CC^{1324I}}\!_P
\]
and pass to the quaternion representation for which
\[
     {\CC^I}\!_{L34}  = -i\Cdot {\CC^I}\!_{L{1324}} \,,\qquad\qquad
     {\CC^I}\!_{L124} = -i\Cdot {\CC^I}\!_{L 4}     \,,\qquad\qquad
     {\CC^I}\!_{L{3}} = -i\Cdot {\CC^I}\!_{L{123}}  \,,
\]
we obtain
\begin{eqnarray*}
     &&{\CC^{mK}}\!_I \Cdot \biggl(\partial_m \psi^I(x) +
     \frac{e\Cdot \AAA_m}{2}\Cdot
     ({\CC^I}\!_{L{21}} + {\CC^I}\!_{L{1324}})\Cdot \psi^L \\
     &&{}+
     \frac{g_\W}{2}\Cdot({\delta^I}\!_P - i\Cdot {\CC^{1324I}}\!_P)\Cdot
     (Z_m\Cdot {\CC^P}\!_{L{1324}} + W^1_m\Cdot {\CC^P}\!_{L{4}} +
     W^2_m\Cdot {\CC^P}\!_{L{123}})\Cdot \psi^L\biggr) =
     \frac{\mc}{2}\Cdot ({\delta^K}\!_L + {\CC^K}\!_{L34})
      \Cdot \psi^L\,.
\end{eqnarray*}
After the substitution of parastrophic matrices
{\renewcommand{\arraystretch}{0.95}
\tighten
\tabcolsep = 1mm
\[
      {\CC^{1324I}}\!_P =i \,\Cdot
      \hbox{\footnotesize
      \begin{tabular}{|cc|cc|}
        \hline
             &\p1  &     &     \\
        \p1  &     &     &     \\
        \hline
             &     &     &\m1  \\
             &     &\m1  &     \\
        \hline
      \end{tabular}
      }
      \,,\quad {\CC^I}\!_{L{1324}} = i \,\Cdot
      \hbox{\footnotesize
      \begin{tabular}{|cc|cc|}
        \hline
             &\p 1 &     &     \\
        \p 1 &     &     &     \\
        \hline
             &     &     &\p 1 \\
             &     &\p 1 &     \\
        \hline
      \end{tabular}
      }
      \,,\quad {\CC^I}\!_{L 4} = i \,\Cdot
      \hbox{\footnotesize
      \begin{tabular}{|cc|cc|}
        \hline
              &     &     &\p 1 \\
              &     &\m 1 &     \\
        \hline
              &\m 1 &     &     \\
         \p 1 &     &     &     \\
        \hline
      \end{tabular}
      }%
     \,,\quad {\CC^I}\!_{L123} = i \,\Cdot
      \hbox{\footnotesize
      \begin{tabular}{|cc|cc|}
        \hline
              &     &\p i &     \\
              &     &     &\m i \\
        \hline
         \m i &     &     &     \\
              &\p i &     &     \\
        \hline
      \end{tabular}
      }%
      \,,
\]
}%
the quantum mechanics equations with respect to the quaternion components of
wave function take the form
{\tighten
\begin{eqnarray*}
     &&||{\CC^{mK}}\!_I ||
     \Cdot \left\{ \partial_m\Cdot
     \hbox{\footnotesize
     $\begin{array}{||l||}
     \Psi^{\scriptscriptstyle 0  }   \\
     \Psi^{\scriptscriptstyle 34 }   \\
     \Psi^{\scriptscriptstyle 123}   \\
     \Psi^{\scriptscriptstyle 124}   \\
     \end{array}$}
     \, + \, \frac{i\Cdot e\Cdot \AAA_m}{2}\Cdot
     \hbox{\footnotesize
     $\begin{array}{||l||}
     \Psi^{\scriptscriptstyle 0  }+\Psi^{\scriptscriptstyle 34 }   \\
     \Psi^{\scriptscriptstyle 0  }+\Psi^{\scriptscriptstyle 34 }   \\
     \Psi^{\scriptscriptstyle 123}+\Psi^{\scriptscriptstyle 124}   \\
     \Psi^{\scriptscriptstyle 123}+\Psi^{\scriptscriptstyle 124}   \\
     \end{array}$}
     \right. \\
     &&\quad{}+
     \left.
     \frac{i\Cdot g_\W}{2}\Cdot
     \left( Z_m\Cdot
     \hbox{\footnotesize
     $\begin{array}{||l||}
     \Psi^{\scriptscriptstyle 0  }+\Psi^{\scriptscriptstyle 34 }   \\
     \Psi^{\scriptscriptstyle 0  }+\Psi^{\scriptscriptstyle 34 }   \\
     \Psi^{\scriptscriptstyle 124}-\Psi^{\scriptscriptstyle 123}   \\
     \Psi^{\scriptscriptstyle 123}-\Psi^{\scriptscriptstyle 124}   \\
     \end{array}$}
      \,+\,  W_m^1\Cdot
     \hbox{\footnotesize
     $\begin{array}{||l||}
     \Psi^{\scriptscriptstyle 124 }-\Psi^{\scriptscriptstyle 123}   \\
     \Psi^{\scriptscriptstyle 124 }-\Psi^{\scriptscriptstyle 123}   \\
     -\Psi^{\scriptscriptstyle 34 }-\Psi^{\scriptscriptstyle 0  }   \\
     \Psi^{\scriptscriptstyle 34  }+\Psi^{\scriptscriptstyle 0  }   \\
     \end{array}$}
      \,+\,  i\Cdot W_m^2\Cdot
     \hbox{\footnotesize
     $\begin{array}{||l||}
     \Psi^{\scriptscriptstyle 123}-\Psi^{\scriptscriptstyle 124}   \\
     \Psi^{\scriptscriptstyle 123}-\Psi^{\scriptscriptstyle 124}   \\
     -\Psi^{\scriptscriptstyle 0 }+\Psi^{\scriptscriptstyle 34 }   \\
     \Psi^{\scriptscriptstyle 0  }+\Psi^{\scriptscriptstyle 34 }   \\
     \end{array}$}
           \right)
     \right\}
     =
     \frac{\mc}{2} \, \Cdot
     \hbox{\footnotesize
     $\begin{array}{||l||}
     \Psi^{\scriptscriptstyle 0  }+\Psi^{\scriptscriptstyle 34 }   \\
     \Psi^{\scriptscriptstyle 0  }+\Psi^{\scriptscriptstyle 34 }   \\
     \Psi^{\scriptscriptstyle 123}+\Psi^{\scriptscriptstyle 124}   \\
     \Psi^{\scriptscriptstyle 123}+\Psi^{\scriptscriptstyle 124}   \\
     \end{array}$} \;.
\end{eqnarray*}}
From here the quantum mechanics equations for the
leptons of the first generation in the electroweak field follow:
\begin{eqnarray*}
    &&i\Cdot \gamma_1^m \left(
    \partial_m + i\Cdot e\Cdot \AAA_m
    \right) e\RR
    = \mc\Cdot e\LL \,,\\
    &&i\Cdot \gamma_2^m \left(\partial_m e\LL +
    i\Cdot e\Cdot \AAA_m\Cdot e\LL +
    i\Cdot g_\W\Cdot Z_m\Cdot e\LL -
    i\Cdot g_\W\Cdot [W_m^1 - i\Cdot W_m^2]\Cdot \nu_e\LL
    \right)
    = \mc\Cdot e\RR \,,\\
    &&i\Cdot \gamma_1^m \Cdot \partial_m \nu_e\RR = 0 \,,\\
    &&i\Cdot \gamma_2^m \left(\partial_m \nu_e\LL -
    i\Cdot g_\W\Cdot Z_m\Cdot \nu_e\LL -
    i\Cdot g_\W\Cdot [W_m^1 + i\Cdot W_m^2]\Cdot e\LL
    \right) = 0 \,.
\end{eqnarray*}
The right and left electrons interact with the electromagnetic field with the
identical coupling constant.  In this approximation, the right electron does
not interact with the weak field in contrast with the left electron and
neutrino.  The right neutrino does not interact with the electroweak field
and therefore it cannot be revealed in the specified interactions.

\subsubsection{The second approximation}

In the previous Section, the variant of the electroweak theory
was considered such that only the left leptons interact the weak field.
In the wake of the Glashow--Weinberg--Salam theory, we shall try
to take into account
the interaction of the right leptons with the weak $Z$-field.
For this propose, the weak group is supplemented by the turn group
about the axis $\varepsilon_{21}$ in the space of leptons $\B{C}_4$.
In other words, we consider that the group of turns
\[
     h = \varepsilon_0\Cdot\cos \varphi_{21} +
     \varepsilon_{21}\Cdot\sin \varphi_{21} \,
\]
is also responsible for the mixed electroweak interaction with $\AAA$ and $Z$
fields, in addition to the turn group about the axis $\varepsilon_{1324}$.
The gauge charge of this interaction with $Z$-field will be written as
$g_i=-g_1/2$. As before, we assume that the angle $\varphi_{21}$ depends only
on the space-time coordinates, and for this subgroup
\[
     g_{21}\Cdot{A^{21}}_m\equiv \frac{e}{2}\Cdot \AAA_m -
     \frac{g_1}{2}\Cdot Z_m \,.
\]

After the calculations similar to the previous ones, we obtain
\begin{mathletters}
\begin{eqnarray}
\label{F20}
    &&i\Cdot \gamma_1^m \left(
    \partial_m + i\Cdot e\Cdot \AAA_m
    - i\Cdot g_1\Cdot Z_m
    \right) e\RR
    = \mc \Cdot e\LL \,,\\
    &&i\Cdot \gamma_2^m \left(\partial_m e\LL +
    i\Cdot e\Cdot \AAA_m\Cdot e\LL +
    i\Cdot [g_\W - g_1]\Cdot Z_m\Cdot e\LL -
    i\Cdot g_\W\Cdot [W_m^1 - i\Cdot W_m^2]\Cdot \nu_e\LL
    \right)
    = \mc\Cdot e\RR \,,\\
    &&i\Cdot \gamma_1^m \left(\partial_m
    - i\Cdot g_1\Cdot Z_m\right) \nu_e\RR = 0 \,,\\
    &&i\Cdot \gamma_2^m \left(\partial_m \nu_e\LL -
    i\Cdot [g_\W + g_1]\Cdot Z_m\Cdot \nu_e\LL -
    i\Cdot g_\W\Cdot [W_m^1 + i\Cdot W_m^2]\Cdot e\LL
    \right)
    = 0 \,.
\end{eqnarray}
\end{mathletters}
In this approximation,
the right and left electrons interact with the electromagnetic field with
the identical coupling constant, the left electron and neutrino interact
with the weak fields $W$ and $Z$, the right electron and neutrino
interact with the weak $Z$-field.

In \cite{art1} we have shown that such a remarkable phenomenon as an
$e$-$\mu$-$\tau$ universality owes its origin to the algebraic equivalency of
the basis vectors $\varepsilon_{21}$, $\varepsilon_{13}$, $\varepsilon_{32}$
in the space of leptons $\B{C}_4$. Recall that these basis vectors are used
for describing three lepton generations. In virtue of the $e$-$\mu$-$\tau$
universality, the equations for the interaction of the muon and $\tau$-lepton
with the electroweak field are similar to ones presented above.

\subsubsection{Comparison with the Glashow--Weinberg--Salam theory}

According to the Glashow--Weinberg--Salam theory, the lagrangian of
electroweak interaction for the leptons of the first generation has the form
(see, for example, \cite{Ryder}):
\begin{eqnarray*}
     {\cal L} =
     i\Cdot \tld{e}\RR\Cdot \gamma_1^m\Cdot \partial_m e\RR +
     i\Cdot \tld{e}\LL\Cdot \gamma_2^m\Cdot \partial_m e\LL +
     i\Cdot \tld{\nu}_e\LL\Cdot \gamma_2^m\Cdot \partial_m \nu_e\LL -
     g\Cdot \sin\theta_\W\Cdot\tld{e}\RR\Cdot
          \gamma_1^m\Cdot e\RR\Cdot \AAA_m -
     g\Cdot \sin\theta_\W\Cdot\tld{e}\LL\Cdot
          \gamma_2^m\Cdot e\LL\Cdot \AAA_m \\
     {}+
     \frac{g}{\cos\theta_\W}\Cdot\sin^2\theta_\W\Cdot
     \tld{e}\RR\Cdot \gamma_1^m\Cdot e\RR\Cdot Z_m -
     \frac{g}{2\Cdot\cos\theta_\W}\Cdot\cos 2\theta_\W\Cdot
     \tld{e}\LL\Cdot \gamma_2^m\Cdot e\LL\Cdot Z_m +
     \frac{g}{2\Cdot\cos\theta_\W}\Cdot
     \tld{\nu}_e\LL\Cdot \gamma_2^m\Cdot \nu_e\LL\Cdot Z_m \\
     {}+
     \frac{g}{2}\Cdot
     \tld{\nu}_e\LL\Cdot \gamma_2^m\Cdot e\LL\Cdot (W^1_m + i\Cdot W^2_m) +
     \frac{g}{2}\Cdot
     \tld{e}\LL\Cdot \gamma_2^m\Cdot \nu_e\LL\Cdot (W^1_m - i\Cdot W^2_m) -
     \mc\Cdot \tld{e}\RR\Cdot e\LL -
     \mc\Cdot \tld{e}\LL\Cdot e\RR \,,
\end{eqnarray*}
where $\theta_\W$ is the Weinberg's angle, $g$ is the standard
coupling constant appropriate the subgroup $SU(2)$ of the Glashow group.
The grouping of addends with identical conjugate vectors allows to write
the quantum mechanics equations for the leptons of the first generation in the
electroweak field in the Glashow--Weinberg--Salam model.
\begin{eqnarray*}
     &&i\Cdot \gamma_1^m \left(\partial_m
     + i\Cdot g\Cdot \sin\theta_\W\Cdot \AAA_m
     - i\Cdot g\Cdot Z_m
     \Cdot \left[\frac{\sin^2\theta_\W}{\cos\theta_\W}\right]
     \right) \Cdot e\RR = \mc\Cdot e\LL \,,\\
     &&i\Cdot \gamma_2^m \left(\partial_m e\LL +
     i\Cdot g\Cdot \sin\theta_\W\Cdot \AAA_m\Cdot e\LL
     + \frac{i\Cdot g\Cdot Z_m}{2} \Cdot \left[\cos\theta_\W -
     \frac{\sin^2\theta_\W}{\cos\theta_\W}\right] \Cdot e\LL
     - \frac{i\Cdot g\Cdot (W^1_m - i\Cdot W^2_m)}{2}\Cdot\nu_e\LL
     \right)
     = \mc\Cdot e\RR \,,\\
     &&i\Cdot \gamma_2^m \left(\partial_m \nu_e\LL
     - \frac{i\Cdot g\Cdot Z_m}{2}
     \Cdot \left[\cos\theta_\W + \frac{\sin^2\theta_\W}{\cos\theta_\W}\right]
     \Cdot\nu_e\LL
     - \frac{i\Cdot g\Cdot (W^1_m + i\Cdot W^2_m)}{2}\Cdot e\LL
     \right) = 0\,.
\end{eqnarray*}
The comparison of these equations with the system (\ref{F20}-d) shows
the following differences of our theory from the Glashow--Weinberg--Salam
theory:
\begin{enumerate}
\item The system of equations is supplemented by the equation for the right
neutrino.

\item The coupling constants of interaction between the leptons, the electron
and the left neutrino, and the weak field differ from
values obtained in the standard model.  However, the concrete
character of these differences depends on the chosen correspondence of
our coupling coefficients $g\W$ and $g_1$ to the parameters $\theta_\W$
and $g$ of the standard model.  For example, if we identify the
expression $({g\Cdot\sin^2\theta_\W})/({2\Cdot\cos\theta_\W})$ with our
coupling coefficient $g_1$, we see that the coupling constant of
interaction between the right electron and the weak $Z$-field in our
consideration is 2 times less than one in the Glashow--Weinberg--Salam model.

\item The right neutrino interacts only with the weak $Z$-field with the same
coupling constant as the right electron.

\end{enumerate}

\section{Conclusions}

We summarize the more important results found in the previous Sections.
\begin{enumerate}
\item The subgroups of turn group of the generalized action space
are identified with the groups of interior symmetries
responsible for interaction.

\item The direct product of turn groups about the
axes $\varepsilon_{21}$ and $\varepsilon_{1324}$ in the Clifford space
$\B{SC}_4$ is responsible for the electromagnetic interaction of leptons.

\item The direct product of turn groups about the axes $\varepsilon_{21}$,
$\varepsilon_{1324}$, $\varepsilon_4$, $\varepsilon_{123}$,
$\varepsilon_{34}$, $\varepsilon_{124}$, $\varepsilon_{3}$ in $\B{SC}_4$ is
responsible for the weak interaction of leptons.

\item The gauge charge is the similarity factor between
interaction subgroup parameters (turn angles) in
the action space and those in the kinematic space.

\item The relativistic quantum mechanics equations for particles in a gauge
field can be derived from the structure equations of the kinematic algebra.
We emphasize that such a derivation have no need of the gauge principle
which is usually applied for the prolongation of free wave equations onto
gauge transformations.

\item The joint description of particles and antiparticles may be found only
if we shall expand the kinematic algebra through the generalized conjugate
space-time $\BT{X}$ (for more details, see \cite{art1}) and the space
$\BT{U}$ of mappings in $\BT{X}$. In doing so, the relations
(\ref{F_rules}-d) should be supplemented by multiplication rules including
basis vectors of the spaces $\BT{X}$ and $\BT{U}$.

\item With the introduction of the generalized space-time, the space-time and
interior space coordinates unite into the coordinates of single vector.  This
union, combined with the union of wave function components, allows to
consider particles in gauge fields from common positions and to count on that
the development of theory proposed by us will result in the construction of
an unified theory of interactions.

\end{enumerate}


\end{document}